\documentclass[aps,prl,twocolumn,superscriptaddress,reprint]{revtex4-1}

\usepackage[ansinew]{inputenc}
\usepackage[T1]{fontenc}
\usepackage{ae,aecompl}
\usepackage[english]{babel}
\usepackage{subfigure,bm,dcolumn,graphicx,hyperref}
\usepackage{amsmath}
\usepackage{epstopdf}
\usepackage{color}
\usepackage[normalem]{ulem}

\begin{document}

\author{Aldo Isidori}
\affiliation{International School for Advanced Studies (SISSA), Via Bonomea 265, I-34136 Trieste, Italy} 
\author{Maja Berovi\'c}
\affiliation{International School for Advanced Studies (SISSA), Via Bonomea 265, I-34136 Trieste, Italy} 
\author{Laura Fanfarillo}
\affiliation{International School for Advanced Studies (SISSA), Via Bonomea 265, I-34136 Trieste, Italy} 
\affiliation{CNR-IOM Democritos, Via Bonomea 265, I-34136 Trieste, Italy}
\author{Luca de' Medici}
\affiliation{Laboratoire de Physique et d'\'Etude des Mat\'eriaux, UMR8213 CNRS/ESPCI/UPMC, Paris, France} 
\author{Michele Fabrizio}
\affiliation{International School for Advanced Studies (SISSA), Via Bonomea 265, I-34136 Trieste, Italy} 
\author{Massimo Capone}
\affiliation{International School for Advanced Studies (SISSA), Via Bonomea 265, I-34136 Trieste, Italy} 
\affiliation{CNR-IOM Democritos, Via Bonomea 265, I-34136 Trieste, Italy}

\title{Charge disproportionation, mixed valence, and Janus effect in multiorbital systems:\\
A tale of two insulators}
\date{\today}

\begin{abstract}
Multiorbital Hubbard models host strongly correlated `Hund's metals' even for interactions much stronger than the bandwidth. We characterize this interaction-resilient metal as a mixed-valence state. In particular it can be pictured as a bridge between two strongly correlated insulators: a high-spin Mott insulator and a charge-disproportionated insulator which is stabilized by a very large Hund's coupling. This picture is confirmed comparing models with negative and positive Hund's coupling for different fillings. Our results provide a characterization of the Hund's metal state and connect its presence with charge disproportionation, which has indeed been observed in chromates and proposed to play a role in iron-based superconductors.
\end{abstract}

\maketitle

The concept of Hund's metal has been introduced~\cite{haule_njp} to rationalize the observation that, in theoretical models of iron-based superconductors, ruthenates, and other materials,  many observables turn out to be remarkably dependent on Hund's coupling $J$ rather than on the Hubbard interaction $U$. In Ref.~\cite{janus} it has been shown that --for integer occupation different from global half-filling-- Mott localization happens in two steps. As a function of Hubbard $U$, the effective mass enhancement first grows, then flattens and remains relatively large in a wide range of interactions~\cite{janus}, while the critical $U$ for Mott localization is pushed to very large values~\cite{luca11}. Therefore we have a wide region of parameters with a strongly correlated metal which is resilient with respect to Mott localization despite a strong Coulomb interaction. This two-faced role of Hund's coupling has been called Janus effect.  Theoretical studies~\cite{werner08,ldmmcchapter,kotliarnrg,kotliarnrgreview,cornaglia} enriched the picture showing: (i) anomalies of the spin response~\cite{hansmann} and finite-temperature spin-freezing at Hund's metal crossover~\cite{werner08,wernerphasediagram}; (ii) orbital decoupling associated with quenching of charge excitations between different orbitals, favoring a strong differentiation in the degree of correlation of different orbitals (orbital selectivity)~\cite{pietra,fanfarillohund,CaponeNM}; (iii) anomalous increase of charge fluctuations~\cite{fanfarillohund}; (iv)  enhancements of nematic orbital differentiation~\cite{fanfarillonematic} and electronic compressibility~\cite{demedicicompressibilita,Villar18}. The latter have been proposed as boosters for superconductivity and can be naturally connected with charge-ordering  instabilities, which have been proposed in iron-based superconductors~\cite{carretta} and chromium perovskites~\cite{Wu14,Cheng1670,PbCrO3_2}. 

Despite the fast advances of our understanding of these phenomena, the very existence of a metallic solution when both $U$ and $J$ are large remains surprising since both interactions are intuitively expected to constrain electron mobility. In this work we address this and related questions in a three-orbital correlated model,  and we link the survival of Hund's metal to the existence of competing insulators. One of these is a charge-disproportionated insulator compatible with the unusual metal-insulator transition observed in $\rm Pb Cr O_3$~\cite{Wu14,Cheng1670}.

We consider the  three-orbital Kanamori model,
\begin{align}
H_\mathrm{int} & = U\sum_a n_{a\uparrow}n_{a\downarrow} 
+ (U-3J)\sum_{a<b, \, \sigma} n_{a\sigma}n_{b\sigma} \nonumber\\
& + (U-2J)\sum_{a\neq b} n_{a\uparrow}n_{b\downarrow}  
+ J\sum_{a\neq b} d^+_{a\uparrow}d^+_{a\downarrow}\,d_{b\downarrow}d_{b\uparrow} \nonumber\\
& - J \sum_{a\neq b} d^+_{a\uparrow}d_{a\downarrow}\,d^+_{b\downarrow}d_{b\uparrow}\label{kana1}\\
& = (U - 3 J) \frac{\hat{n}^2 }{2} 
- J \left( 2\, {\mathbf{S}}^2 + \frac{1}{2} {\mathbf{L}}^2 \right), \label{kana2}
\end{align}
where $\hat{n} = \sum_{a\sigma}d^\dagger_{a\sigma} d_{a\sigma}$ is the local electron number operator, whose expectation value is the average density $\bar{n}$ ($a=1,2,3$ labels the orbital and $\sigma = \uparrow,\downarrow$ the spin components), while  ${\mathbf{S}} = \frac{1}{2} \sum_{a,\,\sigma\sigma'} d^\dagger_{a\sigma} \hat{\bm{\sigma}}_{\sigma\sigma'} d_{a\sigma'}$ and ${\mathbf{L}} = \sum_{ab,\,\sigma} d^\dagger_{a\sigma} \hat{\bm{\ell}}_{ab} d_{b\sigma}$ are the spin and orbital angular momentum operators, with $\hat{\bm{\sigma}}_{\sigma\sigma'}$ the Pauli matrices and $\hat{\bm{\ell}}^{(a)}_{bc} = - i \epsilon_{abc}$ the $O(3)$ group generators.

The form (\ref{kana2}) is very transparent: the first term describes the on-site repulsion with strength $U*=U-3J$ and the second the exchange mechanism responsible for Hund's rules, which favor, for positive $J$, high spin and orbital angular momentum configurations. The form (\ref{kana1})  details all the microscopic interactions. In particular, $U*$ controls the repulsion between electrons in different orbitals with same spin. We study this model in a wider range of parameters with respect to most previous investigations, considering also (i)  the regime  $U* <0$, which is usually discarded because some Coulomb matrix elements are negative, and (ii) both positive and negative $J$. Negative values of $U*$ and $J$ have been proposed in some compounds. 
The model (\ref{kana2}) with $J<0$ has been introduced~\cite{CaponeScience,CaponeRMP,Nomura} for alkali-doped fullerides $A_3 \rm C _{60}$ ($A =\rm K$, Rb, Cs), where  the coupling between electrons and vibrational modes of the $\rm C_{60}$ molecule induces an effective interaction of the same form of the $J$-term but with opposite sign. Since the amplitude of the phonon contribution exceeds  the electronic one, the net exchange coupling is negative~\cite{CaponeScience,CaponeRMP,Nomura}. 
A negative $U*$ has been proposed for rare-earth nickelates~\cite{Subedi} and it can be realised by different screening processes involving, for example, the coupling to a breathing phonon mode or to the other nickel $d$ orbitals and oxygen $p$ orbitals~\cite{Seth}.

We use rotation invariant slave bosons (RISB)~\cite{risb} to find approximate solutions to the model (\ref{kana1}). Technical details about the approach can be found in the supplementary information. Within RISB the lattice enters only through the density of states, which for simplicity we take flat with a width $W$. In order to gain insight on the physics behind the stabilization of Hund's metal, we limit ourselves to paramagnetic solutions without spatial ordering, for which the details of the lattice are not crucial. Broken-symmetry solutions for $U* >0$ have been reported in Ref. \onlinecite{Hoshino3}.
The use of RISB allows us to investigate a wide range of parameters at a moderate cost, describing the development of electronic correlation in terms of the amplitudes (or probabilities) for the local configurations $p_{n,L,S}$, where $n$, $L$, and $S$ are the local occupation, angular momentum, and spin.

\begin{figure}[tb]
        \centering
        \includegraphics[width=\columnwidth]{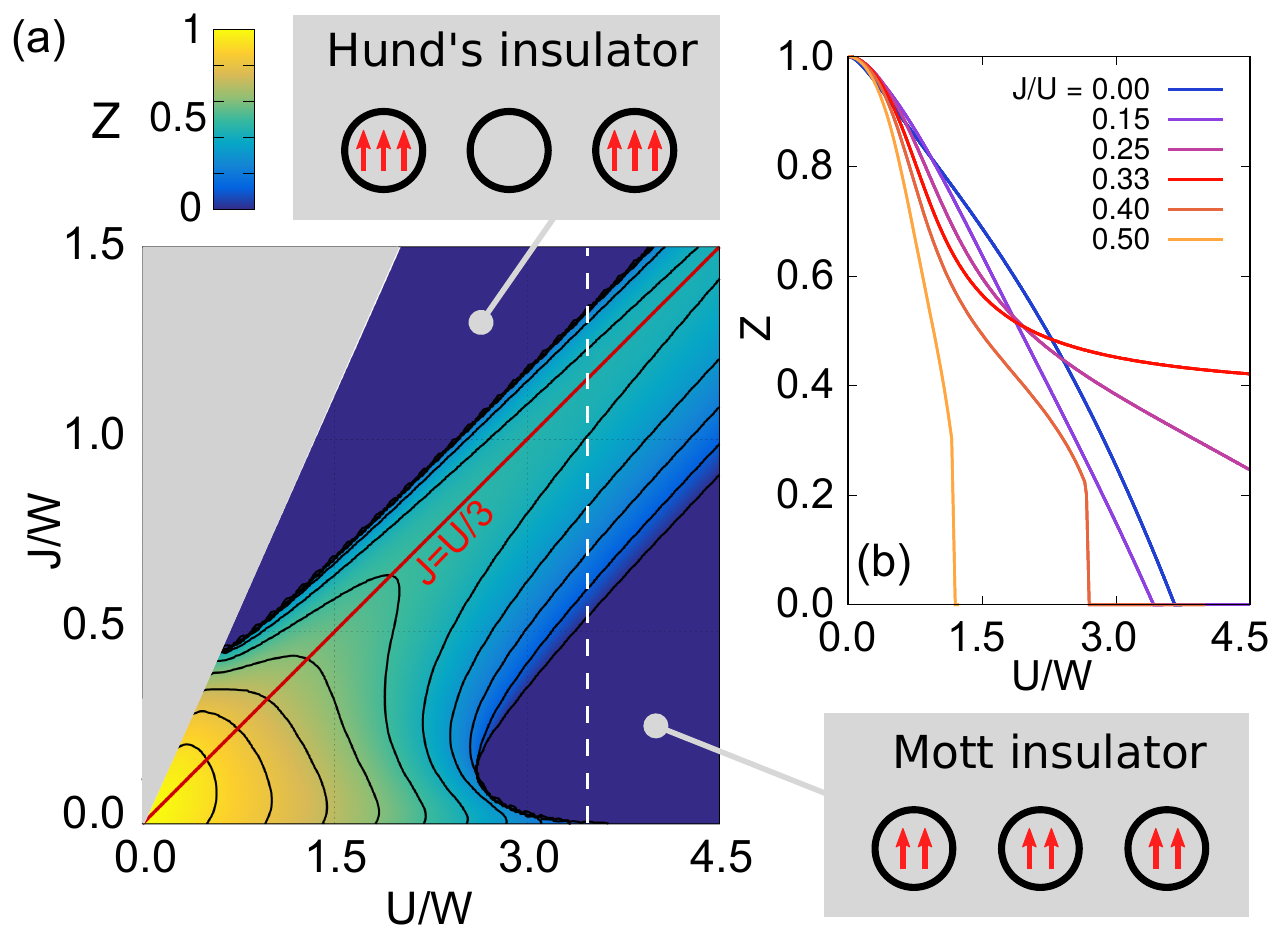}
        \caption{(a) Phase diagram from a color plot of $Z$ for $J>0$ and $\bar{n}=2$. The dark blue region on the right is a Mott insulator (sketch on the right) whereas in the upper left part there are two insulating states: a Hund insulator (sketch on the top) in the dark blue region and an extreme negative charge-transfer insulator in the greyed area. (b): $Z$ plotted along different cuts at fixed $J/U$.} \label{fig:Phase_Diagram_Hund}
\end{figure}

We start from $J>0$ and  $\bar{n}=2$, a paradigmatic case for Hund's metal phenomenology~\cite{janus}. In Fig.~\ref{fig:Phase_Diagram_Hund} we show the quasiparticle weight $Z$, which measures the degree of metallicity. Values close to 1 (light) correspond to a non-interacting metal, while $Z=0$ (dark blue) describes localized electrons of a correlated insulator. In our RISB local mean-field approach $1/Z$ represents also the correlation-driven mass enhancement. A first look at the diagram shows that the metallic region  is stable in a ribbon around the $U*=0$ line, surrounded by two insulating phases. The insulator for positive $U*$ is the standard high-spin Mott insulator (MI). Here charge fluctuations are frozen and each lattice site is occupied by two electrons in the lowest energy multiplet with $S=L=1$. A large $U*$ localizes the electrons and $J$ arranges them in a high-spin configuration. For $J > U/3$ we find instead a charge-disproportionated insulator, where two local configurations with different charge are simultaneously present: empty sites, with a weight of 1/3, and spin-$3/2$ triply occupied sites, with a weight of 2/3~\cite{hugo}. This state is the result of an attractive $U* < 0$ and a large Hund's coupling, favoring configurations with three electrons so as to achieve the highest possible spin. To retain an average density $\bar{n}=2$, one site must remain empty every two triply-occupied sites. We label this state as Hund's insulator (HI). Within our paramagnetic RISB solution empty sites are randomly distributed. If we allow for spatial symmetry breaking, we can expect some ordering.  Perturbation theory for large $J$ is expected to give rise to antiferromagnetic and non-local repulsion of order $t^2/J$, and three-electron hopping terms of order $t^3/J^2$. The competition  can lead to charge ordering or even to metallic behavior, depending on the nature of the lattice. We do not undertake the study of such lattice-dependent phases since our Hund's metal picture would apply also if the pure HI turns into an ordered phase or into a very bad metal.

\begin{figure}[tb]
        \centering
        \includegraphics[width=\columnwidth]{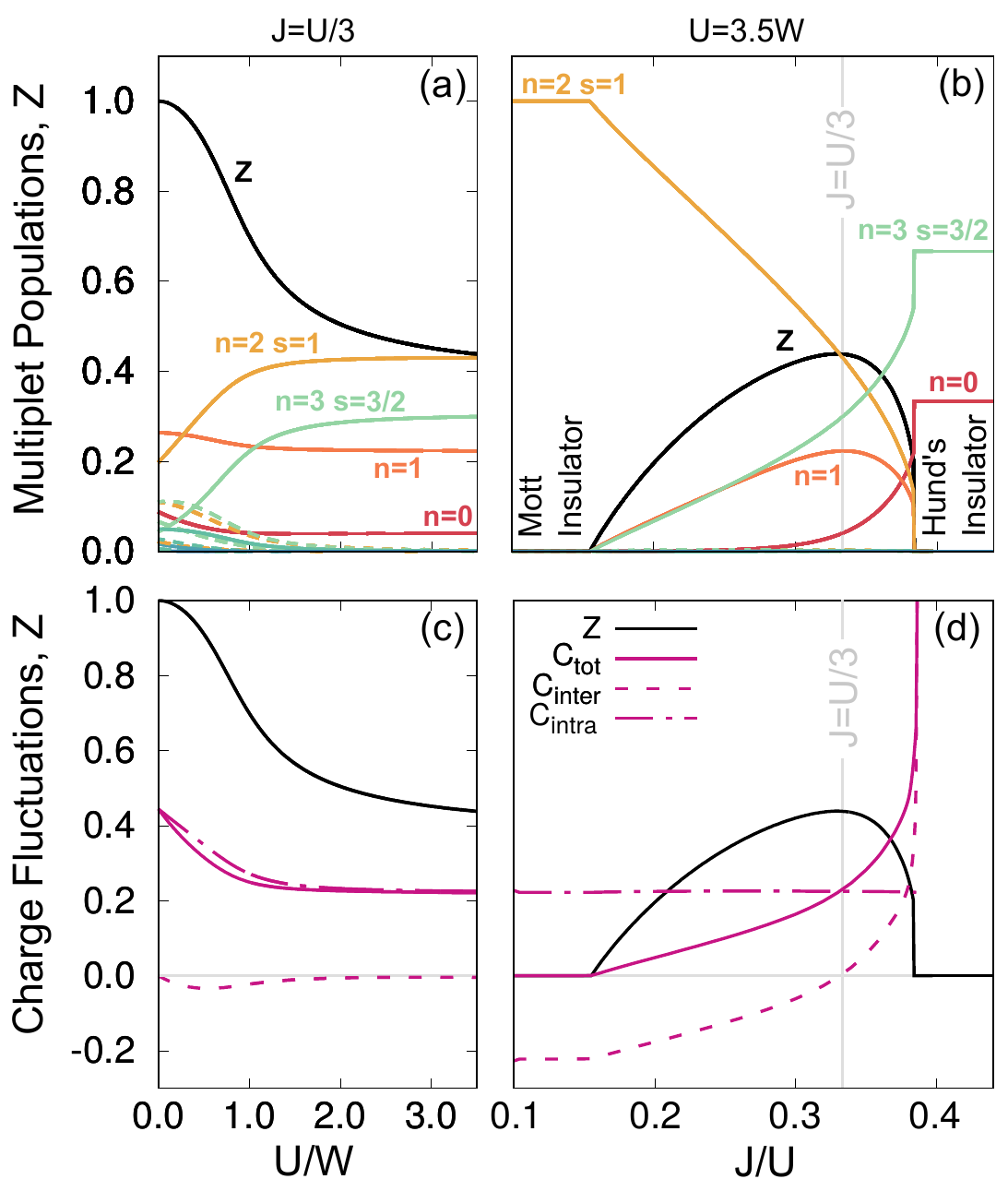}
        \caption{First row: population of the various multiplets: (a) as a function of $U/W$ at $J/U=1/3$; (b) as a function of $J/U$ at fixed $U/W=3.5$. Second row: charge-charge correlations (total, interorbital, and intraorbital); (c) and (d) refer to the same parameters as in (a) and (b), respectively.} \label{fig:Janus_Hund}
\end{figure}

The two insulating solutions have purely atomic character and we can understand their competition in the atomic limit $W=0$. The MI is the ground state for $0< J/U < 1/3$ and the HI for  $1/3 < J/U < 3/4$. The two insulators are degenerate for $J/U=1/3$. Interestingly, on the same line a third insulator with either singly or (high-spin) triply occupied sites, with equal probabilities $p_{1,1,1/2} = p_{3,0,3/2} = 1/2$, is degenerate alongside the MI and HI. Although this state is never the lowest-energy insulator, its presence is relevant for the properties of the metallic regime.  For $ J/U > 3/4 $  the (unphysically) large negative $U*$ favors maximal charge imbalance, so that each site is occupied by either 0 or 6 electrons with weight $p_{0,0,0} = 2/3$ and $p_{6,0,0} = 1/3$ (grey region in the plot).

We now turn to the metallic region separating the MI and HI. Figure~\ref{fig:Phase_Diagram_Hund}~(b) shows cuts of $Z$ at fixed $J/U$. For $J/U$ close to $1/3$ we observe the two-step localization (`Janus effect'), with $Z$ remaining asymptotically finite along the degeneracy line. In Fig.~\ref{fig:Janus_Hund}~(a), instead, we follow the population of each multiplet as a function of $U$, keeping $J/U$ fixed to  1/3. For small $J$ and $U$ (bottom left part of Fig.~\ref{fig:Phase_Diagram_Hund} phase diagram) all atomic multiplets are populated and hopping takes place between any pairs of multiplets that are compatible with Wigner-Eckart theorem. An increase of either $U$ or $J$ has the same net effect of reducing $Z$ by disfavoring some multiplets, thereby reducing the phase space for hopping: an increase of $U$ reduces the population of multiplets with particle number different from the average density (in this case, $\bar{n}=2$), whereas an increase of $J$ tends to reduce the population of multiplets with low spin.  

For weak interactions the reduction is relatively slow, but when we reach the Hund's metal crossover, the differentiation in multiplet occupations suddenly becomes strong: most multiplets have basically zero population, while only a few states survive with a large probability that remains almost constant by further increasing the interaction. These are the high-spin states for each charge sector, which are also the building blocks of the three degenerate insulators. However, this strong phase-space limitation  does not lead to a suppression of $Z$ as one might naively expect.  The reason is that  among the surviving multiplets we have high-spin states whose local occupation differs by one unit, allowing for  hopping processes.  This is only possible  owing to the disproportionated nature of the HI and the other degenerate insulator. 
Exactly on the $U*=0$ line we have finite hopping even when $U/W$ and $J/W$ go to infinity. When we depart from this line, the atomic states are no longer degenerate. However, reasonably close to it, the energy splitting is small. This implies that a relatively small  hopping value is sufficient to overcome the splitting and give rise to a metal. As the ratio $J/U$ moves away from 1/3, the critical $U$ is more and more reduced, as shown in Fig.~\ref{fig:Phase_Diagram_Hund}~(b). The evolution of multiplet occupations along a large $U$ cut connecting the MI and HI (white dashed line of Fig.~\ref{fig:Phase_Diagram_Hund}, fixed $U = 3.5W$) is shown in Fig.~\ref{fig:Janus_Hund}~(b). The main message is that $Z$ is maximized when the populations of the various multiplets become comparable, thereby optimizing the kinetic energy. This observation leads us to identify  the Hund's metal as a mixed-valence state which exists because of the degeneracy (or near degeneracy) between atomic insulators. Along the line  $J/U = 3/4 $ the HI and the fully disproportionated insulator are degenerate. Yet, no hopping process is possible between the atomic states characterizing the two insulators and no metallic solution appears.

In Fig.~\ref{fig:Janus_Hund}~(c) and (d) we plot the total, interorbital, and intraorbital local charge correlations, $ C_{ab} = \langle \hat{n}_a \hat{n}_b \rangle - \langle \hat{n}_a \rangle \langle \hat{n}_b \rangle $, along the same lines of panels (a) and (b), respectively. Panel (c) shows that, inside Hund's metal, interorbital correlations are zero along the $U=3J$ line, reflecting the physics of orbital decoupling~\cite{fanfarillohund}, whereas intraorbital correlations are finite, so that the total charge fluctuations remain finite. More interestingly, panel (d) reveals that intraorbital fluctuations are frozen to a constant value throughout the large $U$ cut connecting the MI and HI. This is due to the complete quench of intraorbital double occupancies. Instead, interorbital correlations remain active and their variation drives the change in the total charge fluctuations. The latter are indeed zero in the MI and become finite when entering the Hund's metal, increasing rapidly as we approach the charge-disproportionated HI (namely for $J > U/3$, when interorbital correlations change sign). The increase of charge correlations preludes to a phase-separation instability. The above phenomenology is consistent with Ref.~\cite{fanfarillohund}, where the behavior of charge fluctuations is understood via the energetics of hopping processes between the allowed multiplets.

\begin{figure}[tb]
        \centering
        \includegraphics[width=\columnwidth]{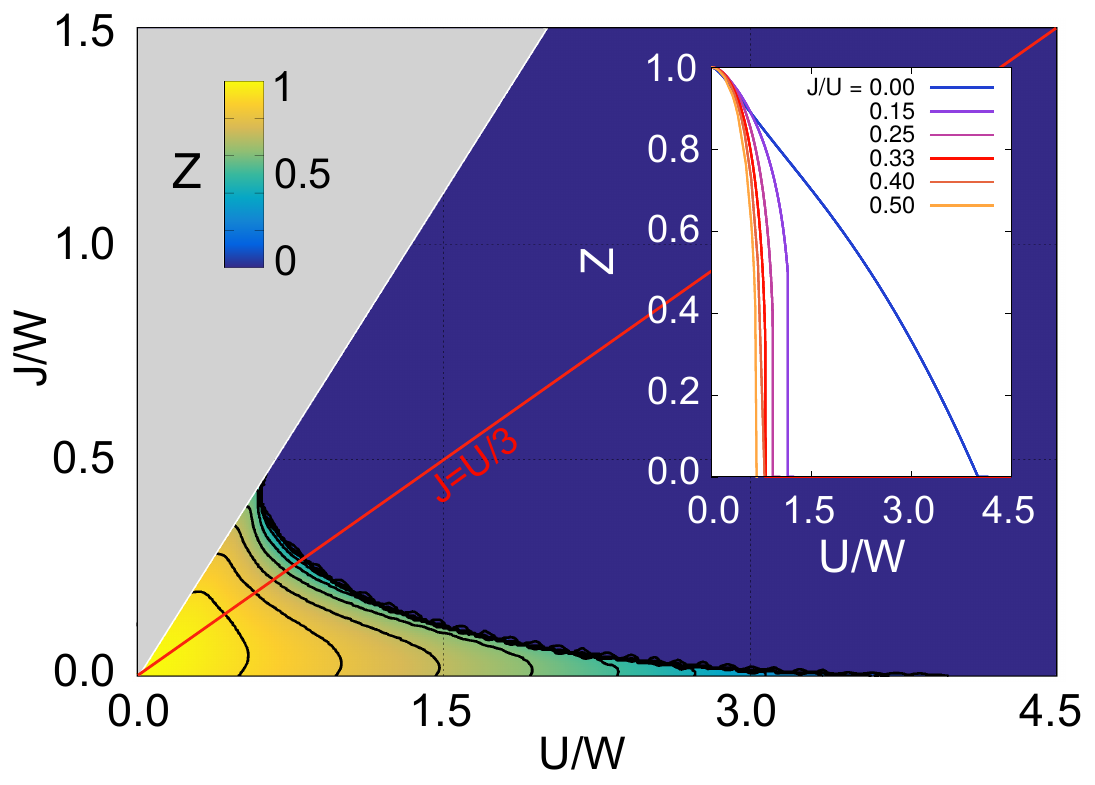}
        \caption{$Z$ of the Hund system at $\bar{n}=3$. The dark blue region is the Mott-Hund insulator, whereas the lighter regions ($Z \neq 0$) denote a metallic state. The inset shows cuts of $Z$ at fixed $J/U$.} \label{fig:Hund_n3}
\end{figure}

Our analysis can be used to interpret experiments on $A \rm Cr O_3$ compounds, where the $3d$ orbitals are populated by two electrons per site. BaCrO$_3$ has been characterized as a Hund's system which becomes insulating because of spin-orbital ordering~\cite{BaCrO3,giovannettibacro3}, while $A =\rm Ca$ is a surprising antiferromagnetic metal with structural anomalies~\cite{CaCrO3} and $A =\rm Sr$ displays phase separation~\cite{SrCrO3}. Most importantly for this manuscript, $\rm Pb Cr O_3$ is experimentally characterized as a charge-disproportionated insulator. References~\cite{Wu14,Cheng1670} propose the same pattern we find for $U < 3J$, whereas Ref.~\cite{PbCrO3_2} invokes a different disproportionation pattern, although with the same net effect of maximizing the total spin. This scenario is compatible with a crossover where the Ba compound has a positive $U*$ which is progressively reduced when moving towards the Ca and Sr systems, becoming eventually negative with Pb. This evolution of the effective interaction can be attributed to electron-phonon coupling, which might further strengthen the charge disproportionation by lowering the lattice symmetry.

At global half-filling, $\bar{n}=3$, the Janus effect does not occur (see Fig.~\ref{fig:Hund_n3}). The key point is that, in this case, the MI and HI coincide and consist in a uniform solution with three electrons per site and $S=3/2$. In other words, $U$ and $J$ do not compete but rather cooperate to create the maximal spin state. Hence, the region of large $U$ and $J$ is simply occupied by this Mott/Hund insulator. All multiplets except the ground state are filtered out by interactions and the lack of degenerate states in the atomic regime completely destroys Hund's metal physics, confirming the crucial role of degenerate insulators in the genesis of the correlation-resilient metal.

The maximal degeneracy found on the $U=3J$ line is specific to the Kanamori interaction, in particular to the ratio $r = 4$ between the coefficients of the ${\mathbf{S}}^2$ and ${\mathbf{L}}^2$ terms in Eq.~(\ref{kana2}).  For $r \neq 4$ --yet still enforcing the spin and orbital rotation invariance-- the degeneracy between the MI and HI is preserved (at a different $J/U$ ratio) and the Hund's metal survives. More complex models, such as the full Coulomb interaction for the whole $d$ manifold, reduce the symmetry of the problem and split the degeneracy between the insulators. However, as long as the different insulating states are reasonably close in energy, we still expect a Hund's metal phenomenology with a finite critical value of $U/W$.

\begin{figure}[tb]
        \centering
        \includegraphics[width=\columnwidth]{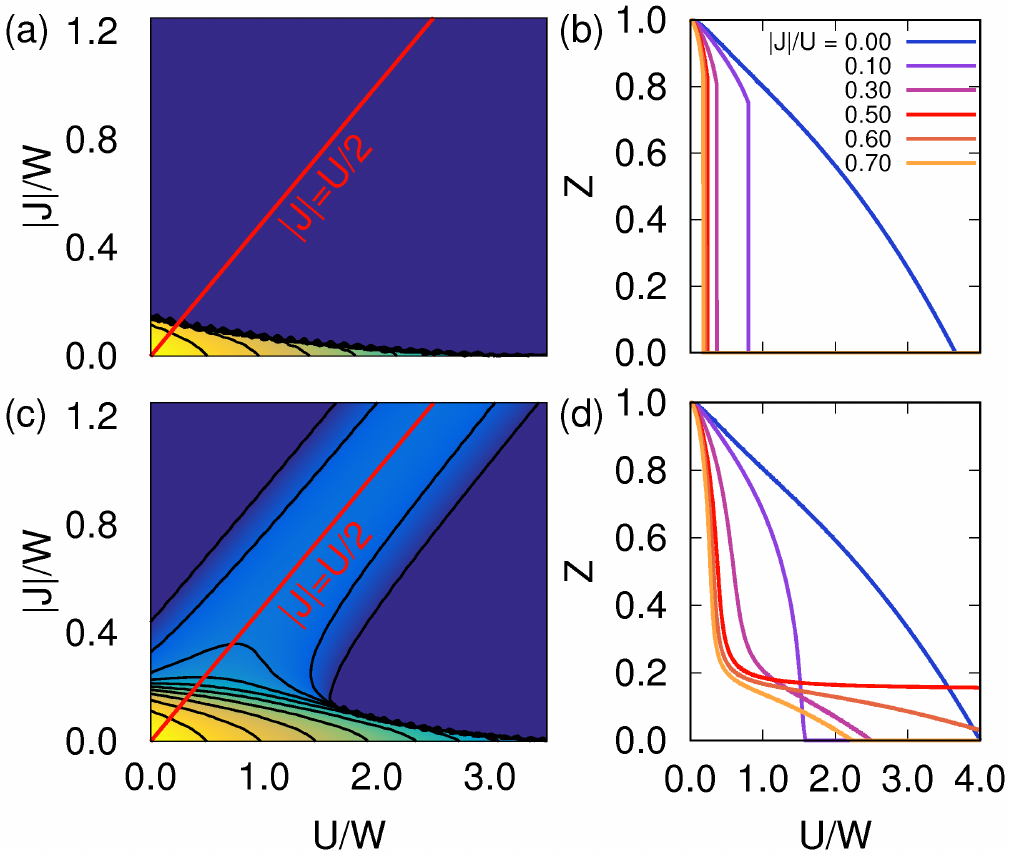}
        \caption{$Z$ in a JT system at $\bar{n}=2$ (a,b) and $\bar{n}=3$ (c,d). For $\bar{n}=3$ and $|J|/U = 1/2$ (degeneracy between Mott and pairing insulators), $Z$ remains asymptotically finite at arbitrarily large $U$. Corresponding cuts at fixed $|J|/U$ are shown in the right panels (b,d).} \label{fig:Janus_JT}
\end{figure}

The link between the persistence of metallic solutions at strong coupling and the degeneracy between atomic insulators can be strengthened by analyzing the model for negative $J$, which has been introduced to study alkali-metal doped fullerides. 
Although realistic estimates for fullerenes imply $U \gg |J|$, leading to superconductivity adjacent to the Mott insulator~\cite{CaponeScience,CaponeRMP,Nomura,WernerC60,risbaldo}, we shall nonetheless study the $J<0$ model in the whole parameter space. We focus on the underlying normal state (metallic or insulating) in order to unveil the attractive counterpart of Hund's metal. In Fig.~\ref{fig:Janus_JT} we show the quasiparticle weight in the paramagnetic state at $\bar{n}=2$ and $\bar{n}=3$.  Notice the strong similarity between the $\bar{n}=3$ JT system and the $\bar{n}=2$ Hund system (cf.\ Fig.~\ref{fig:Phase_Diagram_Hund}). Analyzing the atomic limit for $\bar{n}=3$, also here we find that the MI, characterized by three electrons on each site populating the lowest energy multiplet $L=1$ and $S=1/2$, becomes degenerate with a distinct and competing insulating state at a certain $|J|/U$ ratio, in this case $1/2$. Indeed, for $|J|/U > 1/2$ the energetically favorable insulator is an inhomogeneous state consisting of spin and orbital singlets, where every site is occupied by either two or four electrons, populating the $L=S=0$ states with fractions $p_{2,0,0} = p_{4,0,0} = 1/2$. Charge fluctuations are gapped here by the singlet binding energy. This disproportionated insulator can be seen as a pairing insulator~\cite{Keller,pairinginsulator} formed by an incoherent collection of pairs which are not allowed to condense in a superfluid. 

Like in the $\bar{n}=2$ Hund system, a metallic solution intrudes between the two insulators. The evolution of multiplet occupations mirrors the results of Fig.~\ref{fig:Janus_Hund}.  For small interactions all multiplets are populated. Then, a sharp crossover leads to a highly correlated JT metal~\cite{Zadik15,WernerC60} (in analogy with Hund's metal), where the only populated multiplets are those already present in the insulating states, namely the low-spin states with 2, 3, or 4 electrons per site. The emergent JT metal can be regarded as a mixed-valence state where, due to the degeneracy between insulating states, the systems gains kinetic energy via hopping processes connecting the local configurations of the two insulators. The maximum energy gain and largest $Z$ are reached exactly on the line of degeneracy, where the ground state is precisely halfway between the two insulators, i.e., $p_{3,1,1/2} = p_{2,0,0} + p_{4,0,0} = 1/2$. The similarity between JT and Hund's metals is further strengthened by the `Janus effect', shown in  Fig.~\ref{fig:Janus_JT}~(d). The lower degree of degeneracy of the JT system results in a smaller asymptotic value of $Z$ with respect to the Hund system. For $\bar{n}=2$ we recover a situation similar to the $\bar{n}=3$ Hund case. Here the pairing insulator (lowest spin state) coincides with the MI (minimal charge imbalance), therefore $U$ and $J$ promote the same insulating state. Thus the metallic solution survives only in the bottom left part of the phase diagram and no JT metal emerges.
We finally notice that, despite the common origin of the mixed-valence metals,  in the JT system the charge-transfer gap remains positive for any $J<0$ and does not vanish on the line of degeneracy. 

In this manuscript we have established a strong link between the existence of an interaction-resilient Hund's (or Jahn-Teller) metal and the degeneracy between competing insulators. In the paradigmatic case of a Kanamori model with two electrons in three orbitals, the two insulators are a high-spin MI and a charge-disproportionated HI. The Hund's metal appears as a mixed-valence state which fluctuates between the atomic configurations of the two insulators.  For three electrons in three orbitals the two interactions favor the same insulating solution and, accordingly, Hund's metal physics disappears. A negative Hund's coupling gives rise to an analogous scenario, where for three electrons we find a JT metal~\cite{Zadik15,WernerC60} intruding between two degenerate insulators, while for two electrons the interactions cooperate to form a Mott-pairing insulator. We eventually connect Hund's driven correlations with charge disproportionation. Remarkably, the same disproportionation pattern characterizing the HI has been experimentally observed in PbCrO$_3$.

\begin{acknowledgments}
A.I., M.F., and M.C. acknowledge support from the H2020 Framework Programme, under ERC Advanced GA No. 692670 ``FIRSTORM''. L.F. acknowledges the cost action Nanocohybri CA16218. L.d.M. acknowledges H2020  ERC-StG2016, StrongCoPhy4Energy, GA No. 724177. M.C. also acknowledges financial support from MIUR PRIN 2015 (Prot. 2015C5SEJJ001) and SISSA/CNR project "Superconductivity, Ferroelectricity and Magnetism in bad metals" (Prot. 232/2015). 
\end{acknowledgments}


%


\newpage

\section{Rotation invariant slave bosons: formalism and numerical implementation}

The results presented in the main text are obtained by applying a mean-field approximation scheme, namely the rotation invariant slave boson (RISB) formalism, to the three-orbital Hubbard-Kanamori model, whose local Hamiltonian reads
\begin{align}
H_\mathrm{int} & = U\sum_a n_{a\uparrow}n_{a\downarrow} 
+ (U-3J)\sum_{a<b, \, \sigma} n_{a\sigma}n_{b\sigma} \nonumber\\
& + (U-2J)\sum_{a\neq b} n_{a\uparrow}n_{b\downarrow}  
+ J\sum_{a\neq b} d^+_{a\uparrow}d^+_{a\downarrow}\,d_{b\downarrow}d_{b\uparrow} \nonumber\\
& - J \sum_{a\neq b} d^+_{a\uparrow}d_{a\downarrow}\,d^+_{b\downarrow}d_{b\uparrow} .
\label{kana1}
\end{align}
This method, whose technical details are thoroughly illustrated in Refs.~\cite{risb} and \cite{risbaldo}, provides a reliable description of the Fermi liquid behavior of a strongly correlated multiorbital system at a moderate numerical cost, thereby allowing us to explore the model in a wide range of physical parameters. The rotation invariant nature of the formalism is particularly well suited for investigating multiorbital systems since it allows to treat all inter- and intra-orbital local interactions on the same footing, including non density-density terms such as pair-hopping and spin-exchange (last two terms of Eq.~(\ref{kana1})). This point is crucial in a study, such as the present one, that wants to assess the interplay between the local Coulomb interaction $U$ and Hund's exchange coupling $J$.

In the RISB approach, the local physical Hilbert space of electronic states is mapped onto an enlarged Hilbert space consisting of bosons and auxiliary fermions,
\begin{equation}
 |\Gamma\rangle \; \mapsto \;  
 |\underline{\Gamma}\rangle \equiv \frac{1}{\sqrt{D_\Gamma}}
 \sum_n \phi_{\Gamma n}^\dagger {|\mathrm{vac}\rangle} \otimes |n\rangle_f, 
 \label{eq:phys_ket}
\end{equation}
where $\{|\Gamma\rangle\}$ is a basis set for the physical Hilbert space and $\{|n\rangle_f\}$ are quasiparticle Fock states generated by the auxiliary fermions,
\begin{equation}
 |n\rangle_f \equiv  
 \left(f_1^\dagger\right)^{n_1} \cdots \left(f_M^\dagger\right)^{n_M} 
 {|\mathrm{vac}\rangle}, \quad [n_\alpha = 0,1].
\end{equation} 
Here $\alpha=1,\ldots,M$ labels the local electronic flavors, namely orbital and spin; $M$ is hence twice the number of orbitals. A boson field $\phi_{\Gamma n}$ is introduced for any pair of physical and quasiparticle states that have the same particle number, $N_\Gamma = \sum_\alpha n_\alpha$, as long as one is not interested in allowing for superconductivity (the appropriate modification of the formalism in the presence of superconducting order is detailed in Ref.~\cite{risbaldo}). The normalization factor $D_\Gamma$ appearing in Eq.~(\ref{eq:phys_ket}) represents then the number of quasiparticle states with particle number equal to $N_\Gamma$, 
\begin{equation}
D_\Gamma = \binom{M}{N_\Gamma}.
\end{equation}
The rotational invariance of the formalism resides in the freedom (gauge symmetry) to arbitrarily choose independent basis sets for the physical and quasiparticle Hilbert spaces.  Taking advantage of this freedom, for practical purposes the most convenient choice for the physical basis is given by the set of eigenstates of the local Hamiltonian, $H_\mathrm{int} |\Gamma\rangle\ = E_\Gamma |\Gamma\rangle $, which are often referred to as the atomic multiplets. In order to guarantee a one-to-one mapping between physical states and states of the enlarged Hilbert space, as expressed in Eq.~(\ref{eq:phys_ket}), it is necessary and sufficient to fulfill, as operator identities, the following constraints,
\begin{eqnarray}
 \sum_{\Gamma n} \phi_{\Gamma n}^\dagger \phi_{\Gamma n} &=& 1, \label{eq:const1} \\
 \sum_{\Gamma n n'} \phi_{\Gamma n}^\dagger \phi_{\Gamma n'}
   \langle n'| f_\alpha^\dagger f_{\alpha'} |n \rangle &=& f_\alpha^\dagger f_{\alpha'}. \label{eq:const2}
\end{eqnarray}
The above identities are typically enforced by introducing Lagrange multiplier fields.
Finally, we conclude the setup of the formalism by writing the expression of the physical electron operator and physical observables in terms of the auxiliary fields. The former operator, $\underline{d}_\alpha^\dagger$, must be defined in order to have the following action on the states of the enlarged Hilbert space,
\begin{equation}
 \underline{d}_\alpha^\dagger |\underline{\Gamma} \rangle =
 \sum_{\Gamma'} \langle {\Gamma'}| d_\alpha^\dagger | \Gamma \rangle\, 
 |\underline{\Gamma}' \rangle,
\end{equation}
so that the matrix elements of $\underline{d}_\alpha^\dagger$ in the enlarged Hilbert space are the same of ${d}_\alpha^\dagger$ between corresponding states of the physical Hilbert space. This can be achieved by writing
\begin{equation}
 \underline{d}_\alpha^\dagger = \sum_{\beta} {R}[\phi]_{\alpha\beta}^* f_\beta^\dagger,
\end{equation}
with 
\begin{align}
 {R}[\phi]_{\alpha\beta}^*  = & \sum_{\Gamma \Gamma',n m,\gamma}
 \langle \Gamma'| d_\alpha^\dagger | \Gamma \rangle 
 \phi_{\Gamma' n}^\dagger \phi_{\Gamma m} 
 \langle n| f_\gamma^\dagger |m \rangle \nonumber \\
 & \times \left[\frac{1}{\sqrt{ \hat{\Delta}(1 - \hat{\Delta})} }\right]_{\gamma \beta},\\
 {\Delta}_{\alpha\beta}  = & \sum_{\Gamma n m} \phi_{\Gamma n}^\dagger \phi_{\Gamma m}
 \langle m| f_\alpha^\dagger f_{\beta} |n \rangle.
\end{align}
Analogously, physical observables are also defined such that their matrix elements in the enlarged and physical Hilbert spaces match. In particular, normal local observables, consisting of an equal number of electron creation and annihilation operators, are written in the enlarged Hilbert space as
\begin{equation}
 {\cal O}[\underline{d}^\dagger, \underline{d}] = 
 \sum_{\Gamma \Gamma'} \langle \Gamma'| {\cal O}[{d}^\dagger, {d}] | \Gamma \rangle
 \sum_n \phi_{\Gamma' n}^\dagger \phi_{\Gamma n}. \label{eq:loc_observ}
\end{equation}
Note that in all above expressions we have omitted an implicit site index on both the boson fields and auxiliary fermions, in order to lighten the notation.

The general purpose of the formalism is to disentangle the low-energy nonlocal Fermi liquid behavior, whose coherent quasiparticles are described by the auxiliary fermions, from bosonic collective modes and incoherent high-energy dynamics governed by local interactions, which are now described by the boson fields. In fact, the fermionic Hamiltonian turns out to be quadratic in the auxiliary fermions, with renormalized hopping due to the bosons, and the local interaction is as well quadratic in the bosons. The non-linearity of the problem, of course, does not simply disappear, but it now resides in the presence of Lagrange multiplier fields and boson-dependent hopping amplitudes, which in the RISB mean-field approximation are treated at saddle-point level.


\begin{table*}[tb]
\begin{center}
\begin{tabular}{|c| l c |c|}
\hline
$\displaystyle \phantom{\int}$
$ |\Gamma\rangle $
$\displaystyle \phantom{\int}$   &   $(n,L,S)$   &   $g(n,L,S)$   &   $E(n,L,S)$ \\
\hline
\hline
$\displaystyle \phantom{\int}$
$|0,0,0\rangle$  
$\displaystyle \phantom{\int}$
& $(0,0,0)$  &  $1$  &  $ 0 $ \\
\hline
\hline 
$\displaystyle \phantom{\int}$
$ \frac{1}{\sqrt{2}} \left( \left|\uparrow,0,0 \right\rangle 
+ i \left|0,\uparrow,0 \right\rangle \right)$
$\displaystyle \phantom{\int}$  
& $(1,1,1/2)$  &  $6$  &  $ 0 $ \\
\hline
\hline
$\displaystyle \phantom{\int}$
$ \frac{1}{2} \left( \left|\uparrow\downarrow,0,0\right\rangle 
- \left|0,\uparrow\downarrow,0\right\rangle 
+ i \left|\uparrow,\downarrow,0\right\rangle
- i \left|\downarrow,\uparrow,0\right\rangle \right) $ 
$\displaystyle \phantom{\int}$ 
& $(2,2,0)$  &  $5$  &  $ U - J $ \\
\hline
$\displaystyle \phantom{\int}$
$ \frac{1}{\sqrt{2}} \left( \left|\uparrow,0,\uparrow \right\rangle 
+ i \left|0,\uparrow,\uparrow \right\rangle \right)$ 
$\displaystyle \phantom{\int}$
& $(2,1,1)$  &  $9$  &  $ U - 3 J $ \\
\hline
$\displaystyle \phantom{\int}$
$ \frac{1}{\sqrt{3}} \left( \left|\uparrow\downarrow,0,0\right\rangle
+ \left|0,\uparrow\downarrow,0\right\rangle
+ \left|0,0,\uparrow\downarrow\right\rangle \right) $ 
$\displaystyle \phantom{\int}$
& $(2,0,0)$  &  $1$  &  $ U + 2 J $ \\
\hline
\hline
$\displaystyle \phantom{\int}$
$ \frac{1}{2} \left( \left|\uparrow\downarrow,0,\uparrow\right\rangle 
- \left|0,\uparrow\downarrow,\uparrow\right\rangle 
+ i \left|\uparrow,\downarrow,\uparrow\right\rangle
- i \left|\downarrow,\uparrow,\uparrow\right\rangle \right) $ 
$\displaystyle \phantom{\int}$ 
& $(3,2,1/2)$  &  $10$  &  $ 3 U - 6 J $ \\
\hline
$\displaystyle \phantom{\int}$
$ \frac{1}{2} \left( \left|\uparrow,0,\uparrow\downarrow\right\rangle 
+ i \left|0,\uparrow,\uparrow\downarrow\right\rangle 
+ \left|\uparrow,\uparrow\downarrow,0\right\rangle
+ i \left|\uparrow\downarrow,\uparrow,0\right\rangle \right) $
$\displaystyle \phantom{\int}$  
& $(3,1,1/2)$  &  $6$  &  $ 3 U - 4 J $ \\
\hline
$\displaystyle \phantom{\int}$
$\left|\uparrow, \uparrow, \uparrow \right\rangle$
$\displaystyle \phantom{\int}$  
& $(3,0,3/2)$  &  $4$  &  $ 3 U - 9 J $ \\
\hline
\hline
$\displaystyle \phantom{\int}$
$ \frac{1}{2} \left( \left|\uparrow\downarrow,0,\uparrow\downarrow\right\rangle 
- \left|0,\uparrow\downarrow,\uparrow\downarrow\right\rangle 
+ i \left|\uparrow,\downarrow,\uparrow\downarrow\right\rangle
- i \left|\downarrow,\uparrow,\uparrow\downarrow\right\rangle \right) $ 
$\displaystyle \phantom{\int}$ 
& $(4,2,0)$  &  $5$  &  $ 6 U - 11 J $ \\
\hline
$\displaystyle \phantom{\int}$
$ \frac{1}{\sqrt{2}} \left( \left|\uparrow,\uparrow\downarrow,\uparrow \right\rangle 
+ i \left|\uparrow\downarrow,\uparrow,\uparrow \right\rangle \right)$ 
$\displaystyle \phantom{\int}$ 
& $(4,1,1)$  &  $9$  &  $ 6 U - 13 J $ \\
\hline
$\displaystyle \phantom{\int}$
$ \frac{1}{\sqrt{3}} \left( \left|0,\uparrow\downarrow,\uparrow\downarrow\right\rangle
+ \left|\uparrow\downarrow,0,\uparrow\downarrow\right\rangle
+ \left|\uparrow\downarrow,\uparrow\downarrow,0\right\rangle \right) $ 
$\displaystyle \phantom{\int}$ 
& $(4,0,0)$  &  $1$  &  $ 6 U - 8 J $ \\
\hline
\hline
$\displaystyle \phantom{\int}$
$ \frac{1}{\sqrt{2}} \left( 
\left|\uparrow,\uparrow\downarrow,\uparrow\downarrow \right\rangle 
+ i \left|\uparrow\downarrow,\uparrow,\uparrow\downarrow \right\rangle \right)$
$\displaystyle \phantom{\int}$ 
& $(5,1,1/2)$  &  $6$  &  $ 10 U - 20 J $ \\
\hline
\hline
$\displaystyle \phantom{\int}$
$\left|\uparrow \downarrow,\uparrow \downarrow,\uparrow \downarrow \right\rangle$
$\displaystyle \phantom{\int}$ 
& $(6,0,0)$  &  $1$  &  $ 15 U - 30 J $ \\
\hline
\end{tabular}
\end{center}
\caption{
Electronic eigenstates and eigenvalues $E(n,L,S)$ of the local interaction $H_\mathrm{int}$, Eq.~(\ref{kana1}), together with their degeneracy $g(n,L,S)$. In the first column we have explicitly written, in the Fock basis, a representative component of each energy multiplet. The indicated states are, for each multiplet, the eigenstates of $S_z = \frac{1}{2} \sum_{a=1}^3 ( d^\dagger_{a\uparrow} d_{a\uparrow} - d^\dagger_{a\downarrow} d_{a\downarrow} )$ and $L_3 = i \sum_{\sigma=\uparrow,\downarrow} ( d^\dagger_{2\sigma} d_{1\sigma} - d^\dagger_{1\sigma} d_{2\sigma} )$ with the corresponding largest eigenvalue, namely $S_z = S$ and $L_3 = L$.
}\label{tab:amplitudes}
\end{table*}


Before introducing the mean-field saddle-point approximation, the exact expression of the full Hamiltonian in the enlarged Hilbert space reads
\begin{equation}
 \underline{H} = \underline{H}_\mathrm{kin} + \sum_i \left(\underline{H}_\mathrm{int}[i] 
 + \underline{H}_\mathrm{const}[i] \right), \label{eq:H_aux_full}
\end{equation}
where
\begin{align}
 \underline{H}_\mathrm{kin}  = &  \sum_{ij, \, \alpha\beta} 
 t^{\alpha\beta}_{ij} \underline{d}_{\alpha,i}^\dagger \underline{d}_{\beta,j} \nonumber \\
 = &  \sum_{ij, \, \alpha\beta} {f}_{\alpha,i}^\dagger
 \left( \hat{R}[\phi_i]^\dagger \hat{t}_{ij} \hat{R}[\phi_j] \right)_{\alpha\beta}
 {f}_{\beta,j} \, ,\\[2pt]
 \underline{H}_\mathrm{int}[i]  = &  \, \sum_{\Gamma n} 
 E_\Gamma \phi_{\Gamma n,i}^\dagger \phi_{\Gamma n,i} \, ,\\ 
 \underline{H}_\mathrm{const}[i]  = &  \,\, \lambda_{0,i} 
 \left(\sum_{\Gamma n}  \phi_{\Gamma n,i}^\dagger \phi_{\Gamma n,i} -1 \right) \nonumber \\
 &  +  \sum_{\alpha\beta} \Lambda_{\alpha\beta,i} \left( f_{\alpha,i}^\dagger f_{\beta,i}
 \phantom{ \sum_{\Gamma} } \right. \nonumber \\
 &  -  \left. \sum_{\Gamma n n'} \phi_{\Gamma n,i}^\dagger \phi_{\Gamma n',i}
 \langle n'| f_\alpha^\dagger f_{\beta} |n \rangle \right)  .
\end{align}
Here $\hat{t}_{ij}$ denotes the real-space hopping matrix, whose Fourier transform gives
the band dispersion matrix $\hat{\epsilon}({\bf k})$; $E_\Gamma$ are the eigenvalues of the local interaction Hamiltonian; $\lambda_{0,i}$ and $\Lambda_{\alpha\beta,i}$ are Lagrange multiplier fields associated with the local constraints (\ref{eq:const1}) and (\ref{eq:const2}).
As mentioned earlier, the full Hamiltonian (\ref{eq:H_aux_full}) is now quadratic in the auxiliary fermions. Hence, the latter can be integrated out in a functional integral formulation of the problem, yielding a purely bosonic action that depends only on the slave bosons and Lagrange multiplier fields. Importantly, the original interacting part of the Hamiltonian is now contained in the quadratic (i.e., non-interacting) part of the bosonic action, so that the free boson propagators encompass already all the information about the exact eigenstates of the (original) local interaction. Starting from this bosonic action, a mean-field saddle-point approximation is eventually obtained by replacing all fields with their respective expectation values: 
$\phi_{\Gamma n,i}(\tau) \to \langle \phi_{\Gamma n,i}(\tau) \rangle 
\equiv \varphi_{\Gamma n}$;
$\lambda_{0,i}(\tau) \to \langle \lambda_{0,i}(\tau) \rangle \equiv \lambda_{0}$;
$\Lambda_{\alpha\beta,i}(\tau) \to \langle \Lambda_{\alpha\beta,i}(\tau) \rangle 
\equiv \Lambda_{\alpha\beta}$. The actual numerical value of these saddle-point variables is determined by the minimization of the following free-energy functional, 
\begin{eqnarray}
 \Omega[\varphi, \lambda_{0}, \hat{\Lambda}] & = & 
 - \frac{1}{\beta N} \sum_{\bf k} \mathrm{tr} \ln \left[1 + 
 e^{-\beta \left( \hat{R}[\varphi]^\dagger \hat{\epsilon}({\bf k}) \hat{R}[\varphi] +
 \hat{\Lambda} \right) }\right] \nonumber \\
 & + & \sum_{\Gamma n n'} \varphi_{\Gamma n}^* \left[ 
 E_\Gamma \delta_{n n'} + 
 \lambda_{0} \delta_{n n'}   \phantom{\sum_{\alpha\beta}}  \right. \nonumber \\
 & - & \left. \sum_{\alpha\beta} \Lambda_{\alpha\beta} 
 \langle n'| f_\alpha^\dagger f_{\beta} |n \rangle \right] 
 \varphi_{\Gamma n'}  -  \lambda_{0} , \label{eq:free_en}
\end{eqnarray}
where $N$ is the number of sites and $\beta$ the inverse temperature. Once the minimum stationary point of the above functional has been found, the corresponding saddle-point values of the boson amplitudes can be used to determine, via Eq.~(\ref{eq:loc_observ}), all local static observables such as the average occupation of atomic multiplets, magnetic or orbital order moments, spin and charge static susceptibilities, and so on. Moreover, one can obtain the matrix of quasiparticle spectral weights as
\begin{equation}
\hat{Z}[\varphi] = \hat{R}[\varphi] \hat{R}[\varphi]^\dagger .
\end{equation}

We conclude this section by identifying the relevant (i.e., independent) slave-boson amplitudes needed in the numerical solution of our problem, with local interaction given by Eq.~(\ref{kana1}). In principle, in order to describe a completely generic multi-orbital system (with no superconducting order) using the RISB representation (\ref{eq:phys_ket}), one needs $ \sum_{N_\Gamma = 0}^M D_\Gamma^2 = \binom{2 M}{M} $ independent slave-boson amplitudes. However, this number is significantly reduced if one takes into account all the symmetries of the problem and seeks a solution that does not spontaneously break those symmetries. In the case of our three-orbital Hubbard-Kanamori model with degenerate bands, the system is characterized by a $U(1)_C \otimes {\rm SU}(2)_S \otimes {\rm SO}(3)_L$ symmetry, where $C$, $S$, and $L$ refer to charge, spin, and orbital angular momentum transformations, respectively. In this case, the energies of the atomic multiplets (i.e., the eigenvalues of the local Hamiltonian) depend only on the quantum numbers $n$, $L$, and $S$, denoting the total number of electrons, the total orbital angular momentum, and the total spin, respectively. If we then consider a paramagnetic solution that does not break the full symmetry of the model, as it is the case for all results presented in the main text of our work, it is sufficient to consider only one independent slave-boson amplitude for each distinct eigenvalue $E_\Gamma = E(n,L,S)$. In particular, Wigner-Eckart's theorem imposes that the non-zero slave-boson amplitudes should be of the form
\begin{equation}
\varphi_{\Gamma n} = \langle n | \Gamma \rangle \, \Phi (E_\Gamma)
= \langle n | \Gamma \rangle \, \Phi (n,L,S) . \label{eq:norm_amp}
\end{equation}
With the above normalization, the probability to find the system in a given atomic configuration $|\Gamma \rangle$ is given by
\begin{equation}
P(\Gamma) = \sum_n |\varphi_{\Gamma n}|^2  
= |\Phi (E_\Gamma)|^2 . \label{eq:probab_dist}
\end{equation}
Note that Eq.~(\ref{eq:norm_amp}) ensures that all the $g(E_\Gamma)$ components of a given multiplet with energy $E_\Gamma$ have the same probability $|\Phi (E_\Gamma)|^2$, ruling out any possible spin or orbital ordering. Here, 
\begin{equation}
g(E_\Gamma) = g(n,L,S) = (2L + 1)(2S + 1)
\end{equation}
represents the degeneracy associated with a given eigenvalue $E_\Gamma = E(n,L,S)$. For our problem there are in total 13 distinct eigenvalues of the local Hamiltonian, listed in Table~\ref{tab:amplitudes}, and hence a corresponding number of independent slave-boson amplitudes.

The so-called multiplet populations, discussed in the main text and illustrated in Fig.~2 (a) and (b), are defined as the probabilities to find the system in any of the components of a given multiplet. Hence, their expression in terms of the independent slave-boson amplitudes reads
\begin{equation}
p_{n,L,S} = g(n,L,S) |\Phi (n,L,S)|^2 .
\end{equation}

Regarding the details of the band dispersion $\hat{\epsilon}({\bf k})$, in our work we have chosen to use, for the threefold degenerate bands, a flat density of states of bandwidth $W$. Although a more realistic and material-oriented density of states can be straightforwardly implemented in the present formalism, the choice of a simplified density of states is actually motivated by the idea of identifying those physical effects that are driven by the interplay and competition between the Coulomb and Hund's exchange interaction and not necessarily related to a specific lattice structure. We elaborate more on this point in the next section. We also note that we would have obtained qualitatively identical (and, quantitatively, very similar) results, had we used a semicircular density of states, as it is often done in dynamical mean-field theory.

\end{document}